\documentclass[reprint,aps,prd,twocolumn,superscriptaddress,nofootinbib,amsmath,amssymb]{revtex4-2}
\usepackage{graphicx}
\usepackage{dcolumn}
\usepackage{bm}
\usepackage{color}
\usepackage{booktabs}
\usepackage{tabularx}

\usepackage{orcidlink}
\usepackage{hyperref}
\hypersetup{colorlinks=true,linkcolor=blue,filecolor=blue,urlcolor=blue,citecolor=blue}

\begin{document}
	
	\preprint{APS/123-QED}
	
	\title{Considering lensing effect on gravitational wave signals from black holes in mass gap}
	
	\author{Qiyuan Yang \orcidlink{0009-0007-6444-2026}}
	\affiliation{School of Physics and Technology, Wuhan University, Wuhan 430072, China}
    
	\author{Zhi-Qiang You \orcidlink{0000-0002-3309-415X}}
    \affiliation{Institute for Gravitational Wave Astronomy, Henan Academy of Sciences, Zhengzhou 450046, Henan, China}
    
	\author{Xilong Fan \orcidlink{0000-0002-8174-0128}}
	\email{xilong.fan@whu.edu.cn}
	\affiliation{School of Physics and Technology, Wuhan University, Wuhan 430072, China}

	\date{\today}

	\begin{abstract}
    The pair-instability supernova (PISN) mechanism predicts a mass gap in the black hole population, where no stellar-origin black holes are expected to form. However, several binary black hole (BBH) merger events exhibit component masses that appear to lie within the PISN mass gap. If a gravitational-wave (GW) source is lensed, neglecting lensing effects leads to an underestimation of the luminosity distance and hence the redshift, resulting in an overestimation of the source-frame masses and potentially placing them within the mass-gap region. In this work, we analyze two representative events, GW190521 and GW231123. We establish a direct mapping between the lensing magnification and the fraction of posterior samples with source-frame masses below the mass-gap boundary. Adopting a lower bound of $65\,M_{\odot}$, we find that the magnifications required for $90\%$ of the posterior samples to fall below this boundary are $\mu=444$ for GW231123 and $\mu=39$ for GW190521. At these magnifications, the corresponding source-frame masses of the primary black hole are $52^{+10}_{-10}\,M_{\odot}$ and $42^{+19}_{-14}\,M_{\odot}$, with lensed source redshifts of $2.3^{+0.8}_{-0.5}$ and $2.7^{+1.5}_{-1.2}$, respectively. These results provide a quantitative framework for assessing the lensing hypothesis as a possible explanation for BBH mergers observed within the PISN mass gap, and highlight that the extreme magnifications required for GW231123 may challenge the astrophysical plausibility of simple strong-lensing interpretations.

\end{abstract}

	\maketitle
	
	\section{introduction}\label{sec.1}
	Since the first signal GW150914 of gravitational waves (GWs) from a binary black hole (BBH) merger was detected by LIGO~\cite{2015CQGra..32g4001L, 2016PhRvD..93l2003A, 2016PhRvL.116f1102A, 2016PhRvL.116m1103A}, gravitational wave astronomy has opened an entirely new window into the study of black hole (BH) formation and evolution.
	With the subsequent participation of the Virgo~\cite{2015CQGra..32b4001A} and KAGRA~\cite{2012CQGra..29l4007S, 2013PhRvD..88d3007A, 2019NatAs...3...35K, 2021PTEP.2021eA102A}, the LIGO–Virgo–KAGRA (LVK) network has detected hundreds of BBH merger events up to the latest O4 observing run~\cite{2019PhRvX...9c1040A, 2021PhRvX..11b1053A, 2023PhRvX..13d1039A, 2025arXiv250818081T}.
	
	Among these detections, a subset of sources exhibit unusually large component masses, which challenge our current understanding of stellar evolution and black hole (BH) formation channels. GW190521 is particularly remarkable~\cite{2020PhRvL.125j1102A, 2020ApJ...900L..13A}. Detected during the second observing run (O2), it was the most massive BBH merger observed at that time, with component masses of approximately $m_1 \approx 98\,M_{\odot}$ and $m_2 \approx 57\,M_{\odot}$~\cite{2024PhRvD.109b2001A}. This makes it the first event with one BH mass falling within the so-called pair-instability supernova (PISN) mass gap~\cite{1964ApJS....9..201F, 2002ApJ...567..532H, 2010ApJ...724..341H}. The PISN theory predicts that when the helium-core mass of a star lies between the lower boundary of about $45\sim 65\,M_{\odot}$ and the upper boundary of about $120\sim 150\,M_{\odot}$~\cite{2016A&A...594A..97B, 2017ApJ...836..244W, 2019ApJ...887...53F, 2020ApJ...889...75L, 2020A&A...640L..18M, 2020MNRAS.493.4333R, 2025arXiv251018867R, 2025arXiv250904637A, 2025arXiv250909123A, 2025arXiv250904151T}, the star undergoes a pair-instability explosion that completely disrupts it, leaving no compact remnant. Therefore, standard stellar evolution models do not expect the formation of BHs within this mass range. GW190521 is thus regarded as the first BBH event with at least one component located inside the PISN mass gap~\cite{2020ApJ...904L..26F, 2021MNRAS.502L..40F}. 
	More recently, another high-mass event, GW231123, was reported during the latest fourth observing run (O4), with component masses of approximately $m_1 \approx 137\,M_{\odot}$ and $m_2 \approx 101\,M_{\odot}$~\cite{2025ApJ...993L..25A}. It represents the most massive BBH merger event detected so far. Such extreme component masses again far exceed the lower boundary of the PISN mass gap, making GW231123 another potential candidate within this regime. The recurrence of such high-mass events raises a key question: is the upper mass limit of stellar-origin BHs truly as predicted by current theory, or could other astrophysical or observational effects be responsible for the apparent overestimation of masses? The observation of BBH mergers located within the PISN mass gap clearly challenges this traditional picture of BH formation~\cite{2025arXiv250819208M}.
	
	To explain these high-mass merger events, several possible formation channels have been proposed. One explanation suggests that these high-mass systems may originate from primordial black holes, which form in the early Universe through the collapse of enhanced curvature perturbations rather than stellar evolution~\cite{2021PhRvL.126e1101D, 2022PDU....3801111C, 2022PhLB..82937040C, 2022PhRvD.105h3526F, 2025arXiv250809965D, 2025PhRvD.112h1306Y}. Another scenario is that of hierarchical mergers, in which smaller black holes repeatedly merge to gradually form more massive remnants~\cite{2021MNRAS.502.2049L, 2021MNRAS.505..339M, 2022ApJ...941....4A, 2024ApJ...977..220A, 2025arXiv250908298L, 2025arXiv250717551L, 2025arXiv251014363P, 2025arXiv250910609P, 2025arXiv250904151T}. However, the interpretation of BBH mergers occurring within the PISN mass gap remains under debate~\cite{2017PhRvD..96l3523A, 2024arXiv240408416R, 2021NatAs...5..749G, 2021ApJ...918L..31M, 2022ApJ...935L..26F, 2024A&A...688A.148T, 2025ApJ...994L..37K}. 
    In addition, another compelling explanation involves gravitational lensing~\cite{2020arXiv200208821B, 2025arXiv250821262S, 2025arXiv251107186C, 2025arXiv251216916C, 2026PhRvD.113d2006F}. If the GW signal produced by a BBH merger is strongly lensed by a galaxy or galaxy cluster located between the source and the observer, the observed strain amplitude will be magnified~\cite{2018MNRAS.480.3842O, 2018MNRAS.476.2220L, 2020MNRAS.495.3727R, 2020MNRAS.495.1666R}. If the lensing magnification effect is not taken into account, the inferred luminosity distance will be underestimated, leading to an underestimation of the redshift~\cite{2017PhRvD..95d4011D, 2018PhRvD..97b3012N, 2020MNRAS.495.3740P}. Since the true source-frame masses of the BHs are related to the detector-frame masses through the redshift, an underestimated redshift will result in overestimated source-frame masses, causing them to fall into the PISN mass gap. This effect can naturally explain some of the observed massive BBH events in the mass gap without invoking extreme formation scenarios.

     In this work, we investigate the relationship between lensing magnification and the inferred source-frame black hole masses for the events GW190521 and, in particular, GW231123. Specifically, we rescale the posterior luminosity distances of these events over a range of magnification factors and derive the corresponding source-frame masses under the lensing scenario. This enables us to assess whether lensing effects can shift the inferred source-frame masses across the PISN mass-gap boundary. Larger magnifications correspond to larger inferred luminosity distances and hence to smaller source-frame masses, potentially allowing these systems to move out of the mass-gap region. After adopting a lower boundary for the mass gap, we determine the magnification required for a given fraction of posterior samples of the source-frame mass to fall below this threshold, and present the corresponding lensed source-frame masses and redshifts. Based on these results, we model the lens, calculate the optical depth, and derive the relation between the lens redshift and lens mass within the singular isothermal sphere (SIS) model. We also estimate the time delay between multiple images produced by the lens. Our results indicate that GW190521 and GW231123 are unlikely to be explained by simple galaxy-scale strong-lensing scenarios. Nevertheless, this work demonstrates that gravitational lensing may play an important role in parameter inference when interpreting high-mass binary black hole mergers.
    
    In the following sections, we first introduce gravitational lensing in Sec.~\ref{sec.2}. In Sec.~\ref{sec.3}, we investigate how gravitational lensing affects gravitational-wave waveforms and our analysis method. In Sec.~\ref{sec.4}, we present the results for the source-frame masses under lensed scenarios, with particular emphasis on GW231123. Finally, a summary and discussion of this work are provided in Sec.~\ref{sec.5}.
	Throughout this paper, we take the speed of light $c=2.998\times 10^8 \mathrm{m}\cdot\mathrm{s}^{-1}$, the gravitational constant $G=6.67\times 10^{-11}\mathrm{m^3}\cdot\mathrm{kg}^{-1}\cdot\mathrm{s^{-2}}$ and the $\Lambda$CDM cosmology model described in Ref.~\cite{2020A&A...641A...6P}, which is also coded in \textbf{Astropy}\footnote{http://www.astropy.org}~\citep{2022ApJ...935..167A} with assumptions of  $H_{0}=67.66\mathrm{km}\cdot\mathrm{s}^{-1}\cdot\mathrm{Mpc}^{-1}$ and $\Omega_{\mathrm{m}}=0.30966$.

    \section{gravitational lensing}\label{sec.2}

    Gravitational lensing arises from massive objects, whose gravitational fields can bend the propagation of electromagnetic or gravitational waves. The deflection is described by the lens equation:
    \begin{equation}\label{eq.1}
    \boldsymbol{\beta} = \boldsymbol{\theta} - \boldsymbol{\alpha}(\boldsymbol{\theta}),
    \end{equation}
    where $\boldsymbol{\alpha}$ is the deflection angle, $\boldsymbol{\theta}$ denotes the angular position of the image, and $\boldsymbol{\beta}$ represents the angular position of the source. Gravitational lensing can also lead to a magnification of the observed flux. To compute the magnification factor $\mu$, it is necessary to adopt a model for the mass distribution of the lens. For galaxy-scale lenses, a commonly used model is the singular isothermal sphere (SIS). For an SIS lens, when $y<1$, two images are formed with magnifications $\mu_+$ and $\mu_-$ given by:
	\begin{equation}\label{eq.2}
		\mu_{\pm}=1\pm \frac{1}{y} \, ,
	\end{equation}
    where $y$ is the dimensionless source position defined by $y=\beta / \theta_E$. 
    Here $\theta_E$ is the Einstein angular radius, which sets the characteristic angular scale of the lensing system and is given by:
    \begin{equation}\label{eq.3}
    \theta_E = 4\pi \left(\frac{\sigma_v}{c}\right)^2 \frac{D_{ls}}{D_s},
    \end{equation} 
    where $\sigma_v$ is the velocity dispersion of the lens, and $D_s$ and $D_{ls}$ are the angular diameter distance from the observer to the source and from the lens to the source, respectively. And the lens mass $M_l$ is defined as the projected mass enclosed within the Einstein radius, which serves as a characteristic mass scale of the lens:
    \begin{equation}\label{eq.4}
    M_l = \frac{c^2D_l D_s \theta_E^2}{4G D_{ls}},
    \end{equation}
    where $D_l$ is the angular diameter distance from the observer to the lens. Due to the lensing potential, a time delay arises between the multiple images produced by the lens. 
    The relative time delay between the two images is given by:
    \begin{equation}\label{eq.5}
		\Delta t_d=\frac{8 G M_l(1+z_l)y}{c^3} \, .
	\end{equation}

     Given a cosmological model, one can calculate $D_l$, $D_s$ and $D_{ls}$ from the redshift of lens $z_l$ and source $z_s$. 
     For an SIS lens, once the geometrical configuration of the lens system is specified, the magnification is determined by the lens potential and the source position, and conversely, the magnification constrains the combination of these quantities.

    \section{lensing effect on GW}\label{sec.3}

    For an unlensed BBH merger event, the GW waveform $h^{ul}$ can be written as: 
	\begin{equation}\label{eq.6}
		h^{ul}=F_{+}\,h_{+}+F_{\times}\,h_{\times} \,,
	\end{equation} 
	where $h_{+}$, $h_{\times}$ are the two polarization modes of GW and $F_{+}$, $F_{\times}$ are the antenna pattern functions. $h^{ul}$ is characterized by 15 parameters, which can be divided into two types: 9 intrinsic parameters including BHs masses $m_{1,2}$, spin magnitudes $a_{1,2}$, four angles for spin orientation $\theta_1, \theta_2, \phi_{12}, \phi_{\rm{JL}}$, the phase of waveform $\phi$, and another 6 extrinsic parameters, which include right-ascension $\rm{RA}$, declination $\rm{DEC}$, and the luminosity distance $d_L$, the polarization angle $\psi$, the angle between line of sight and the total angular momentum $\theta_{\rm{JN}}$ of the GW source, and the time at which the GW merger signal reaches the geocenter $t_{\rm{geo}}$.

	For the lensed GW, the waveform will be modified by the lensing object. In frequency domain, the relation between a lensed $h^{l}(f)$ and an unlensed GW waveform $ h^{ul}(f)$ is given by~\citep{1992grle.book.....S, 1999PThPS.133..137N, 2003ApJ...595.1039T}:
	
	\begin{equation}\label{eq.7}
		h^{l}(f)=F(f)\cdot h^{ul}(f) \,,
	\end{equation} 
	where $F(f)$ is the amplification factor and is given by the diffraction integral:
	\begin{equation}\label{eq.8}
		F(f)=\frac{D_s\xi_0^2 (1+z_l )}{D_lD_{ls}}\frac{f}{i}\int e^{2\pi ift_d(\mathbf{x,y})}  d^2\mathbf{x} \,,
	\end{equation}
	where $D_l$, $D_s$ and $D_{ls}$ are the angular diameter distances to the lens, source and from the lens to the source, respectively. $z_l$ is the redshift of lens, $\xi_0$ is the normalization constant, also represents the physical Einstein radius in the lens plane. $t_d$ is the time delay relative to the straight path in unlensed propagation, determined by the dimensionless position $\mathbf{x}$ in lens plane and $\mathbf{y}$ in source plane. In actual observations, we only care about the relative time delays between different images, so when calculating $t_d$, we can add an additional phase term to set its minimum value to zero.

	If the lens object is a galaxy or galaxy cluster, its physical scale is typically much larger than the wavelength of GW. Therefore, the calculation of the amplification factor can be treated under the geometric optics approximation. For a lensing system that can produce $N$ multiple images, the total amplification factor is~\citep{1999PThPS.133..137N}:
	\begin{equation}\label{eq.9}
		F(f) = \sum_{i=1}^{N}\sqrt{|\mu_i|} \cdot e^{-i \pi \left(2 f \Delta t_{d_i} + n_i \right)} \,,        
	\end{equation}
	and then the lensed waveform in the frequency domain can be written as:
	\begin{equation}\label{eq.10}
		h^l(f) =  \sum_{i=1}^{N}\sqrt{|\mu_i|} \cdot e^{-i \pi \left(2 f \Delta t_{d_i} + n_i \right)}\cdot h^{ul}(f) \,,
	\end{equation}
	where $\mu_i$ is the magnification of the $i$-th image, $\Delta t_{d_i}$ is the time delay of the $i$-th image relative to the first one (that means $\Delta t_{d_1}=0$), and $n_i=0, \frac{1}{2}, 1$ is called Morse index, when the position of the $i$-th image is a minimum, saddle, and maximum point of $t_d(\mathbf{x,y})$, respectively.
	
	Note that the sign in the exponential term of $F(f)$ in this paper differs from Eq.~(15) in Ref.~\citep{1999PThPS.133..137N} , because we adopt a Fourier transform convention that is consistent with the one commonly used in GW analyses. Specifically, the inverse Fourier transform convention we adopt is given by:
	\begin{equation}\label{eq.11}
		h(t) = \int_{-\infty}^{+\infty} \tilde{h}(f)  e^{2\pi i f t}  df \,,
	\end{equation}
	then in time domain, the lensed waveform of the $i$-th image $h^l_i(t)$ is:
	\begin{equation}\label{eq.12}
		h^l_i(t) = \sqrt{|\mu_i|} \cdot e^{-i\pi n_i} \cdot h^{ul}(t-\Delta t_{d_i}) \,.
	\end{equation}
	This choice of notation ensures consistency with the convention of GW time--frequency transformation. Moreover, the term $h(t-\Delta t_{d_i})$ guarantees that an image with a larger time delay arrives $\Delta t_{d_i}$ seconds later than the first one, in agreement with actual observational situations.

    Since the time delay of the first image $\Delta t_{d_1}=0$ and it is the minimum image of the lens (that means $n_1=0$), for a SIS lens, the lensed GW waveforms $h^l_1(f)$ and $h^l_2(f)$ of the two images are:
	\begin{equation}\label{eq.13}
		h^l_1(f) = \sqrt{|\mu_+|} \cdot h^{ul}(f) \,,
	\end{equation}
	\begin{equation}\label{eq.14}
		h^l_2(f) = \sqrt{|\mu_-|} \cdot e^{-i \pi \left(2 f \Delta t_d + n_2 \right)}\cdot h^{ul}(f) \,.
	\end{equation}

    \begin{figure*} 
		\centering
		\includegraphics[width=\textwidth]{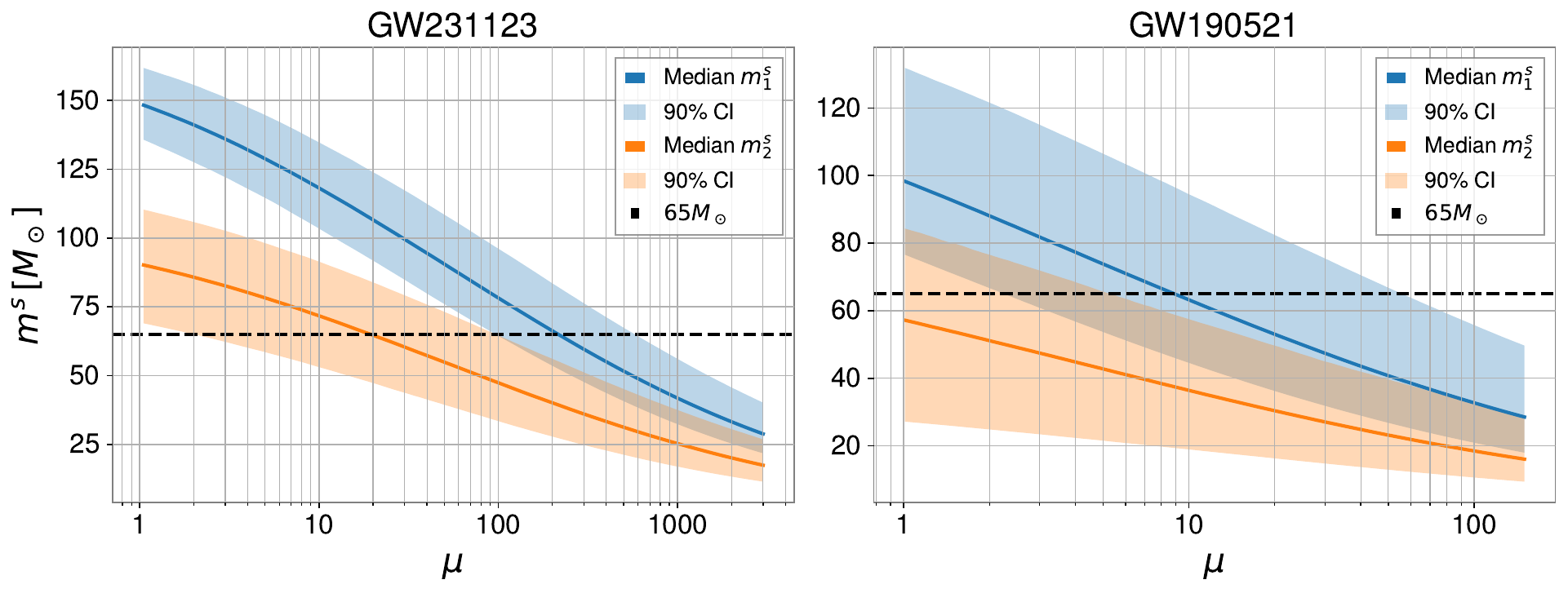} 
		\caption{The relation between the source-frame masses inferred from posterior samples and the lensing magnification $\mu$, left panel for GW231123 and right panel for GW190521. The blue (orange) solid curves represent the median values of $m_1^s$ ($m_2^s$), and the shaded regions indicate the $90\%$ credible intervals (CI). The horizontal dashed black line marks the adopted lower boundary of the mass gap at $m = 65 M_\odot$.}
		\label{fig.1}
	\end{figure*}
    
	Usually, we can obtain the posterior distributions of detector frame masses of each component $m_{1,2}^d$ and the luminosity distance of GW source $d_L$ by Bayesian parameter estimation. And the inferred source-frame masses $m_{1,2}^s$ are constrained by the mass–redshift relation:
	\begin{equation}\label{eq.15}
		m_{1,2}^s = \frac{m_{1,2}^d}{1+z^{}} \,,
	\end{equation} 
	where $z$ is the redshift of GW source. Given a cosmological model, one can calculate $z$ from $d_L$, and then calculate $m^s_{1,2}$. Since the amplitude of GW is inversely proportional to the luminosity distance, then through Eq.~(\ref{eq.13}) we can find that the magnification $\mu$ of a lensed waveform is degenerate with the luminosity distance. For a given GW signal, the luminosity distances with and without lensing satisfy the following relation:
	\begin{equation}\label{eq.16}
		d_L=\frac{d_L^l}{\sqrt{\mu}} \,,
	\end{equation} 
	here $d_L^l$ denotes the true luminosity distance when considering the lensing effect and $d_L$ is the luminosity distance without lensing. In other words, the lensing effect alters our estimation of the luminosity distance. If a GW signal is lensed, the luminosity distance of GW source inferred under the unlensed scenario will be smaller than the true value, which leads to a smaller redshift $z$. So for a lensed GW event, according to Eq.~(\ref{eq.15}), using an unlensed template will lead to an underestimated luminosity distance, which in turn results in overestimated source-frame masses. Such a biased inference may incorrectly place the event within the mass-gap region, like GW190521 and GW231123. Therefore, properly accounting for the lensing effect is essential for a reliable interpretation of the source-frame properties of such events, motivating the analysis presented in this work.

    For these two events, we perform our analysis using the publicly available posterior samples from \textbf{GWOSC}\footnote{https://gwosc.org}~\citep{2025arXiv250818079T}, with the waveform family IMRPhenomXPHM. To investigate whether lensing can shift the inferred source-frame masses out of the mass gap, we adopt a lower boundary of $65\,M_\odot$ and consider a range of magnifications to compute the corresponding source-frame masses. Following Eq.~(\ref{eq.16}), we first rescale the posterior samples of luminosity distance by different magnification factors to obtain the lensed luminosity distance $d_L^l$. This is then converted into the corresponding lensed-redshift $z_s^l(d_L^l)$, from which the lensing-corrected source-frame masses are derived using Eq.~(\ref{eq.15}). We then compute, for each magnification, the fraction of posterior samples with source-frame masses below $65\,M_\odot$. We define $F(m_1^s < 65\,M_\odot \mid \mu_p) = p/100$, i.e., the fraction for which the source-frame mass of the primary BH is below $65\,M_\odot$ at a given magnification $\mu_p$. We further identify three characteristic magnifications, $\mu_{10}$, $\mu_{50}$, and $\mu_{90}$, corresponding to the cases where $10\%$, $50\%$, and $90\%$ of the posterior samples fall below the mass-gap boundary, respectively. Then we model the lens using a SIS profile. Using the inferred redshift $z_s^l$, we compute the corresponding optical depth of lensing by galaxies based on the formula given in Appendix~A.1 of Ref.~\citep{2018arXiv180707062H}. To infer the lens properties, we assume that the lens redshift lies within $0.01 < z_l < z_s^l(d_L^l)$, and consider only the positive-parity image ($\mu_+$), fixing the source position parameter to $\beta = 0.1$. Using Eqs.~(\ref{eq.2}), (\ref{eq.4}), and (\ref{eq.5}), we derive the corresponding lens mass and the relative time delay between the two images as functions of the lens redshift. The results are shown in the next section.

    \begin{figure*} 
		\centering
		\includegraphics[width=\textwidth]{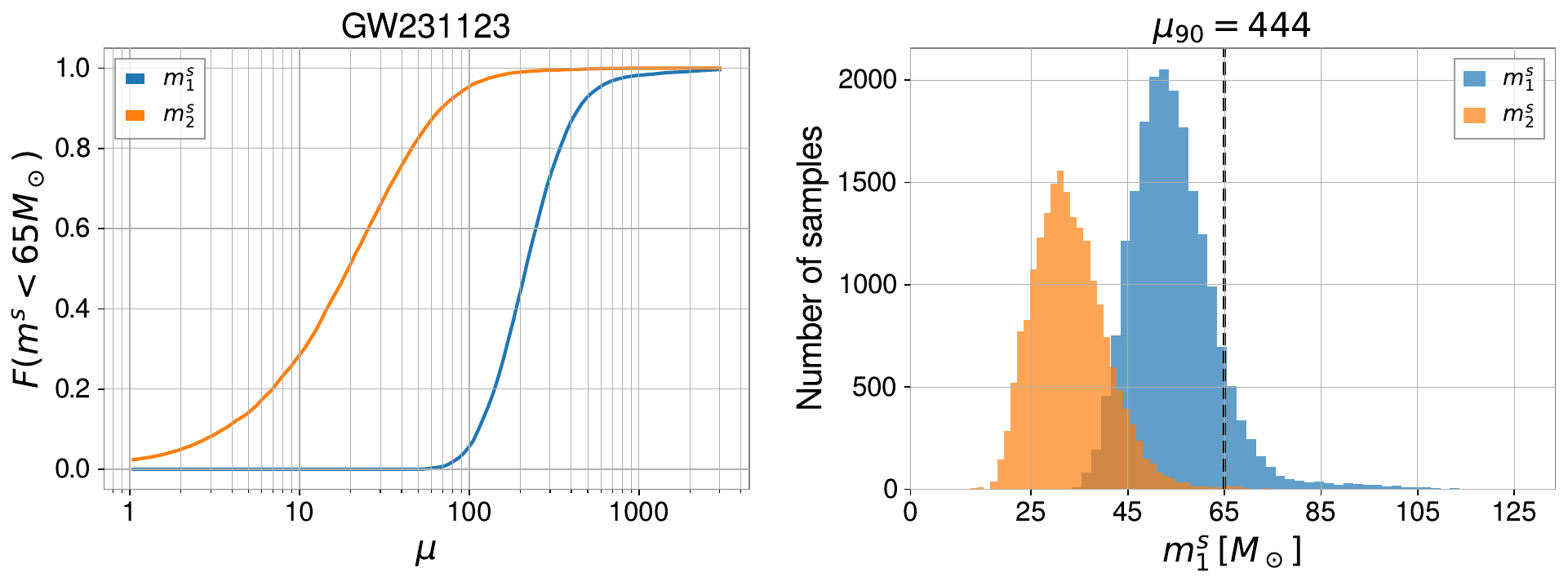} 
		\caption{Left panel: Fraction of posterior samples with source-frame masses below 65$M_\odot$ as a function of the lensing magnification $\mu$ for GW231123. The blue and orange curves correspond to $m_1^s$ and $m_2^s$, respectively. Right panel: Posterior distributions of the source-frame masses $m_1^s$ and $m_2^s$ under a fixed magnification $\mu = 444$. The horizontal dashed black line marks the lower edge of the pair-instability mass gap at $65\,M_\odot$. At this magnification, 90\% of the posterior samples for $m_1^s$ fall below the adopted lower boundary of the mass gap.}
		\label{fig.2}
	\end{figure*}

	\section{result}\label{sec.4}
	Whether or not the lensing effect is taken into account can significantly influence the inferred parameters of a GW source, potentially leading to the appearance of mass gap events like GW190521 and GW231123. In this section, we assess how lensing effect impacts the key parameters and under what conditions it can account for these unusual events.

    Fig.~\ref{fig.1} provides a direct mapping between the lensing magnification and the source-frame mass for the two events. The left panel is for GW231123 and the right panel for GW190521.
    It shows how the posterior distributions of the source-frame masses of the gravitational-wave event vary with magnification. The solid blue and orange curves represent the medians of the posterior distributions for $m_1^s$ and $m_2^s$, respectively, while the shaded regions indicate the 90\% credible intervals (CI). The horizontal black dashed line indicates the adopted lower boundary of the mass gap, $m_g = 65\,M_\odot$.
    When the magnification $\mu = 1$, no lensing effect is present. In this case, the inferred source-frame masses and luminosity distance for both events correspond to the posterior distributions obtained using unlensed waveform templates. As the magnification increases, the source-frame masses inferred from the posterior samples decrease accordingly. For GW231123, the median source-frame mass of the primary BH falls below the adopted mass-gap boundary of $65\,M_\odot$ when $\mu > 216$. For GW190521, this occurs at $\mu > 9$.

    \begin{figure*} 
		\centering
		\includegraphics[width=\textwidth]{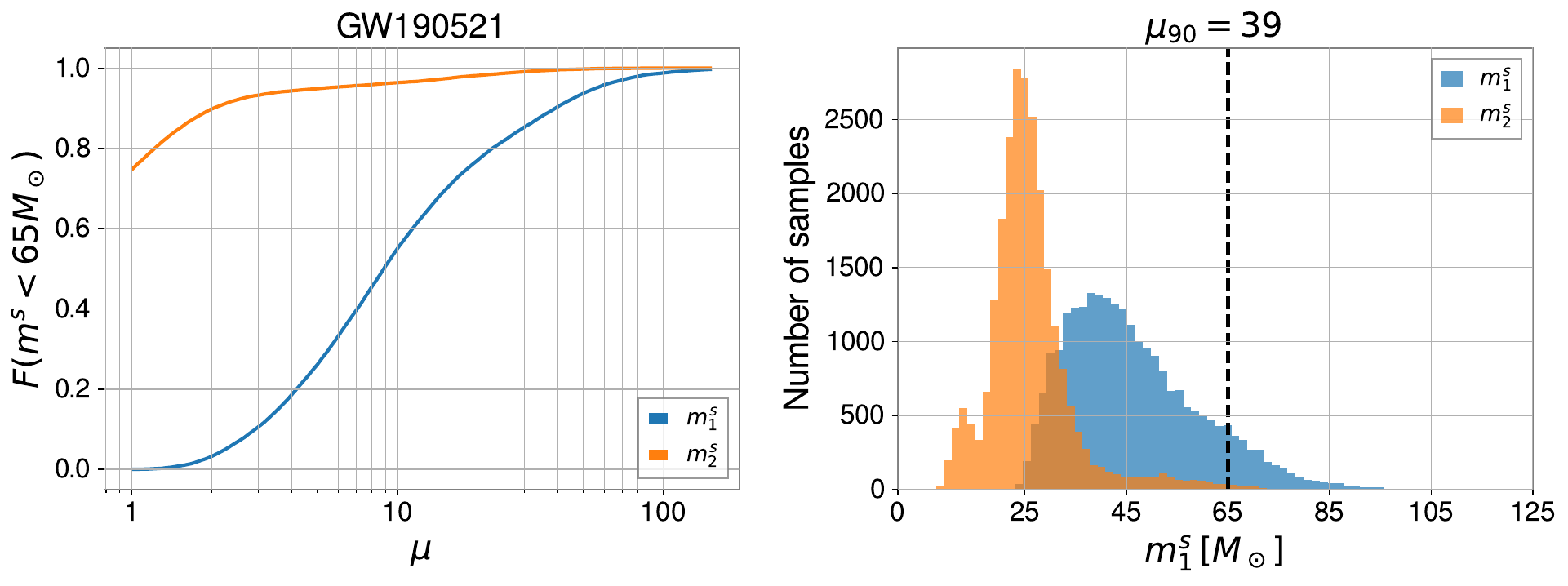} 
		\caption{Same as Fig.~\ref{fig.2}, but for GW190521. The right panel shows the posterior distributions of the source-frame masses $m_1^s$ and $m_2^s$ at a fixed magnification $\mu = 39$. At this magnification, 90\% of the posterior samples for $m_1^s$ fall below the adopted lower boundary of the mass gap.}
		\label{fig.3}
    \end{figure*}

    As shown in the left panels of Fig.~\ref{fig.2} and Fig.~\ref{fig.3}, we present the fraction $F(m_1^s < 65\,M_\odot \mid \mu)$ as a function of the magnification $\mu$ for GW231123 and GW190521, respectively. 
    For GW190521, we obtain $\mu_{10}=3$, $\mu_{50}=9$, and $\mu_{90}=39$. At these magnifications, considering only the posterior samples that satisfy
    $m_1^s<65\,M_\odot$, the medians and 90\% credible intervals 
    of the lensing-corrected source-frame masses of the primary black hole (with the corresponding values for the secondary black hole given in parentheses) are 
    $61^{+3}_{-6}\,(53^{+7}_{-8})\,M_\odot$, 
    $56^{+8}_{-12}\,(38^{+10}_{-9})\,M_\odot$, and 
    $42^{+19}_{-14}\,(25^{+11}_{-8})\,M_\odot$, respectively.  
As shown in the right panel of Fig.~\ref{fig.3}, when the magnification is $\mu_{90}=39$, the median of the source-frame mass distribution for the primary black hole, restricted to the portion lying below the mass gap threshold is $42\,M_\odot$.
    
    The corresponding lensed redshifts are 
    $1.4^{+0.3}_{-0.2}$, 
    $1.8^{+0.6}_{-0.5}$, and 
    $2.7^{+1.5}_{-1.2}$, 
    respectively. The possibility of a strong-lensing origin for GW190521 has been previously discussed in Ref.~\citep{2020ApJ...900L..13A}, which inferred a broad posterior distribution for the magnification, $\mu = 6^{+28}_{-4}$, together with corresponding source-frame masses $m_1^s = 55^{+9}_{-22}\,M_\odot$, $m_2^s=44^{+14}_{-19}\,M_\odot$ and redshift $z = 1.8^{+1.7}_{-0.6}$ . Compared to the more detailed joint inference performed in that work, we adopt a simplified approach based on posterior rescaling, in which the source-frame parameters are evaluated at selected magnifications. To facilitate a comparison, we consider representative magnifications spanning the range reported in Ref.~\citep{2020ApJ...900L..13A}. At $\mu = 2$, we obtain $m_1^s = 63^{+2}_{-4}\,M_\odot$, $m_2^s = 58^{+5}_{-6}\,M_\odot$, and $z = 1.2^{+0.2}_{-0.2}$. At $\mu = 6$, the corresponding values are $m_1^s = 58^{+6}_{-10}\,M_\odot$, $m_2^s = 42^{+10}_{-9}\,M_\odot$, and $z = 1.6^{+0.4}_{-0.4}$. At $\mu = 34$, we obtain $m_1^s = 44^{+18}_{-14}\,M_\odot$, $m_2^s = 26^{+11}_{-8}\,M_\odot$, and $z = 2.5^{+1.4}_{-1.1}$. The corresponding fractions of posterior samples with $m_1^s < 65\,M_\odot$ are $3\%$, $33\%$, and $88\%$ for $\mu = 2$, $6$, and $34$, respectively. Overall, we find that the inferred source-frame masses and redshifts at representative magnifications fall within the broad ranges reported in Ref.~\citep{2020ApJ...900L..13A} for the $m_{\max}=65\,M_\odot$ case. In particular, for $\mu \sim 2$--$34$, our results yield $m_1^s \sim 30$--$65\,M_\odot$ and $z \sim 1$--$3.9$, which are consistent with the parameter scales inferred in that work. This indicates that our posterior-rescaling approach captures the overall behavior of the lensing interpretation, although it does not attempt to reproduce the full joint posterior over magnification and source parameters.

    For GW231123, we obtain $\mu_{10}=114$, $\mu_{50}=216$, and $\mu_{90}=444$. At these magnifications, considering only the posterior samples that satisfy $m_1^s<65\,M_\odot$,
    the lensing-corrected source-frame masses of the primary black hole (with the corresponding values for the secondary black hole given in parentheses) are 
    $62^{+2}_{-6}\,(36^{+10}_{-8})\,M_\odot$, 
    $59^{+5}_{-8}\,(35^{+10}_{-9})\,M_\odot$, and 
    $52^{+10}_{-10}\,(31^{+11}_{-10})\,M_\odot$, with corresponding lensed redshifts 
    $1.8^{+0.3}_{-0.2}$, 
    $1.9^{+0.5}_{-0.3}$, and 
    $2.3^{+0.8}_{-0.5}$, 
    respectively. We also show, in the right panels of Fig.~\ref{fig.2}, the posterior distributions of the source-frame masses at the magnification $\mu_{90}=444$. In this case, $90\%$ of the posterior samples for the primary black hole have masses below $65\,M_\odot$. Such a large magnification is expected to be rare in galaxy-scale strong lensing, as it typically requires a highly fine-tuned source position close to the caustic. Therefore, although this magnification would shift most posterior samples below the mass-gap boundary, the astrophysical probability of realizing such a configuration is likely to be very low. We further compute the optical depth corresponding to the source redshift at $\mu_{90}$, obtaining $7^{+4}_{-2} \times 10^{-4}$. Since the optical depth increases with redshift, even at the highest lensed redshift corresponding to $\mu_{90}$, it remains very small.

\begin{figure} 
		\centering
		\includegraphics[width=\linewidth]{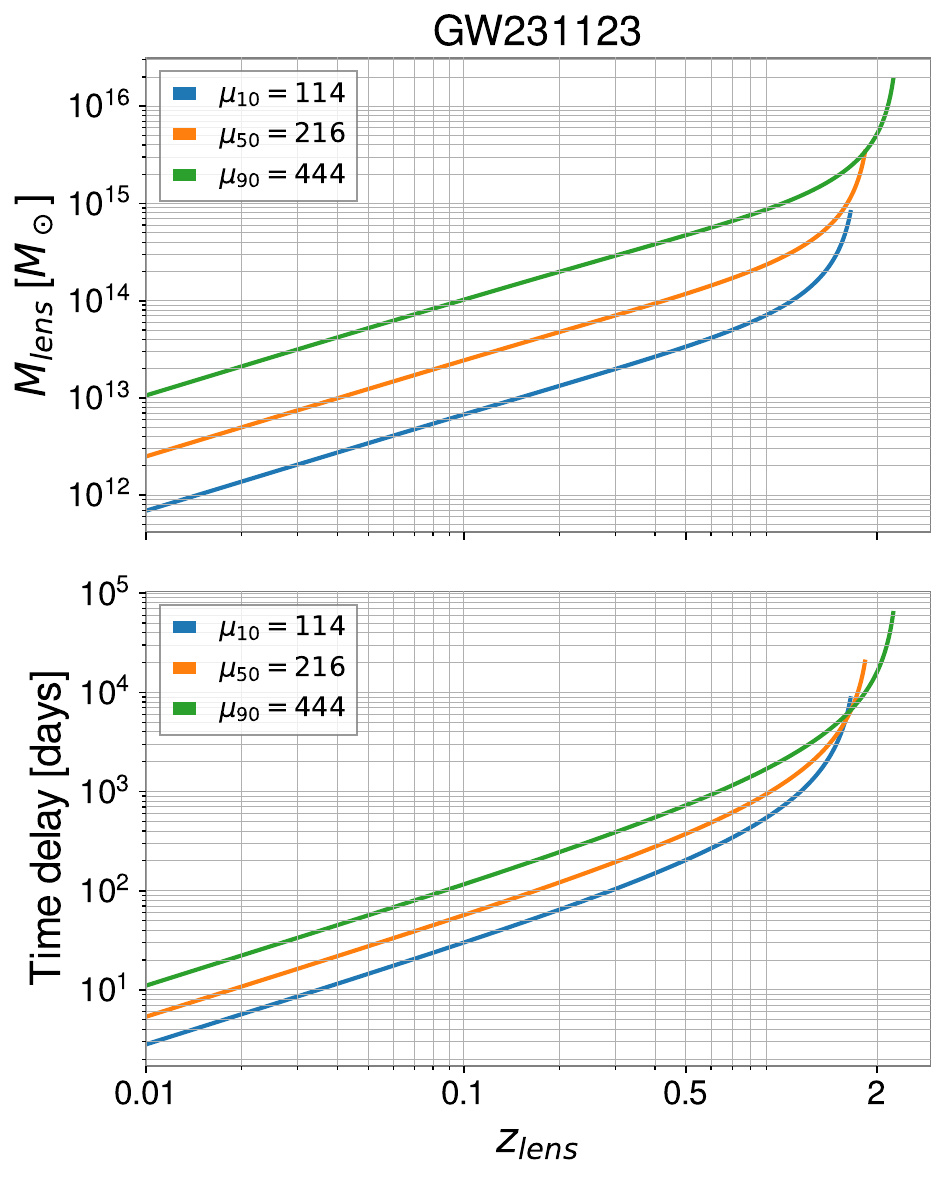} 
		\caption{Upper panel: The relation between the lens redshift and the lens mass required for GW231123 to be shifted out of the mass-gap regime under the lensing hypothesis. The blue, orange, and green curves correspond to the cases with magnifications $\mu_{10}$, $\mu_{50}$, and $\mu_{90}$, respectively. For each magnification, the source redshift is fixed to the median of the corresponding lensed-redshift posterior distribution. Larger magnifications correspond to larger lensed luminosity distances and hence higher inferred source redshifts, allowing for a wider range of possible lens redshifts and requiring more massive lenses. Lower panel: The corresponding time delay between multiple images as a function of the lens redshift for the same set of magnifications. Larger magnifications generally lead to larger lens masses and hence longer time delays.}
		\label{fig.4}
	\end{figure}

\begin{figure} 
		\centering
		\includegraphics[width=\linewidth]{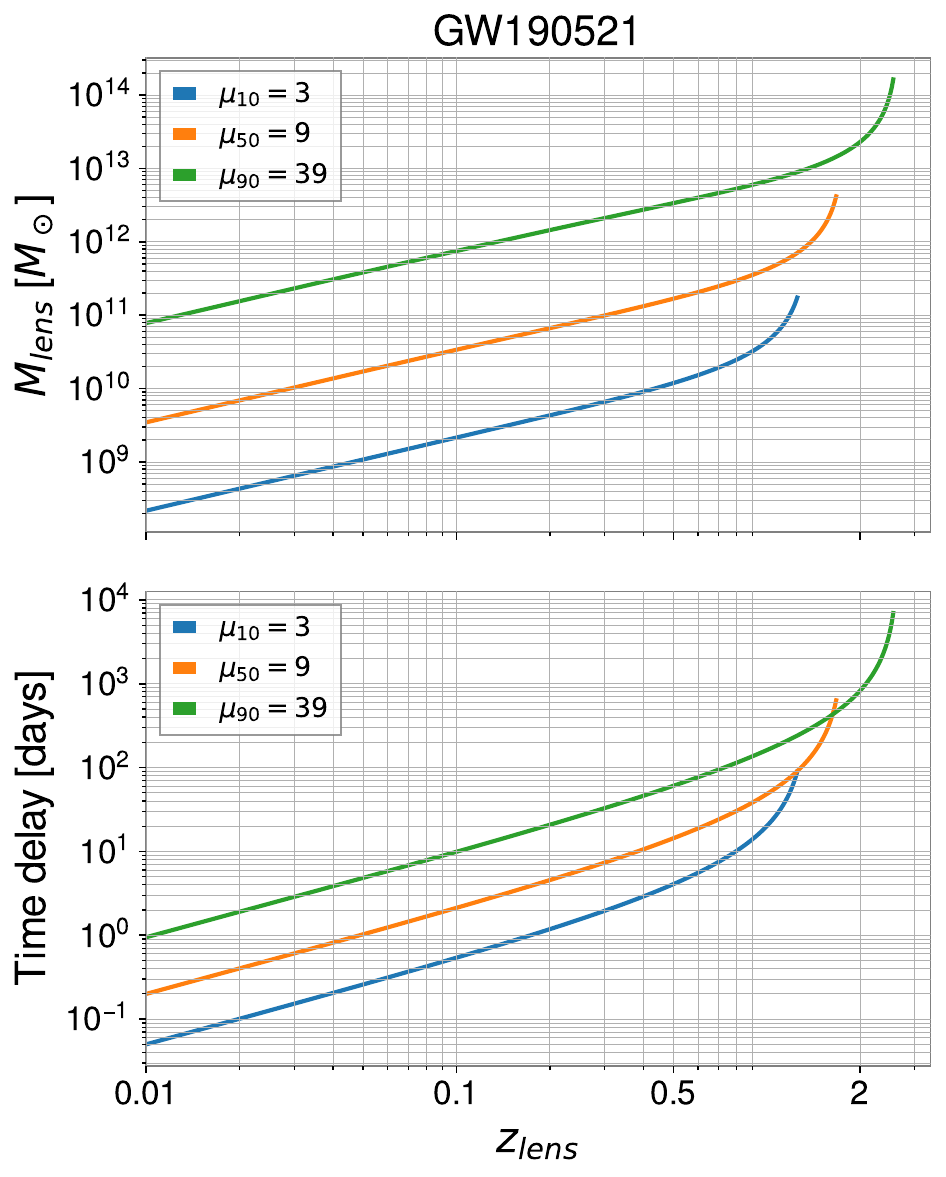} 
		\caption{Same as Fig.~\ref{fig.4}, but for GW190521, showing the relation between lens mass and lens redshift (upper panel) and the corresponding time delays (lower panel) under different magnifications.} 
		\label{fig.5}
	\end{figure}

    After determining the required magnifications, we place constraints on the lens mass using the SIS model. Fig.~\ref{fig.4} and Fig.~\ref{fig.5} show the relation between the lens redshift and the corresponding lens mass (upper panels), as well as the associated time delays between two images as functions of the lens redshift (lower panels), for GW231123 and GW190521, respectively. The blue, orange, and green curves correspond to the cases with magnifications $\mu_{10}$, $\mu_{50}$, and $\mu_{90}$, respectively. At each magnification, we fix the source redshift $z_s^l$ to the median of the corresponding lensed-redshift posterior distribution. For a given magnification, we find that a higher lens redshift requires a more massive lens. Moreover, larger magnifications generally correspond to more massive lenses. We note that the upper limits of the lens redshift differ among the three cases. Since the lens redshift is required to be smaller than the source redshift, larger magnifications lead to higher lensed source redshifts, thereby allowing a wider range of possible lens redshifts. The lower panels show the corresponding time delays between multiple images. Consistent with the behavior of the lens mass, larger magnifications typically result in longer time delays, reflecting the increasing mass scale of the lens.

    To further illustrate the lens properties required to produce such magnifications, we consider a representative lens configuration by fixing the lens redshift to $z_l = z_s^l/2$. Under this assumption, we estimate the corresponding lens masses for the three characteristic magnifications. For GW190521, the inferred lens masses (time delays) are approximately $2\times10^{10}\,M_\odot$ (7 days), $3\times10^{11}\,M_\odot$ (37 days), and $9\times10^{12}\,M_\odot$ (265 days) for $\mu = 3$, $9$, and $39$, respectively. For GW231123, the corresponding values are $7\times10^{13}\,M_\odot$ (534 days), $3\times10^{14}\,M_\odot$ (1075 days), and $10^{15}\,M_\odot$ (2706 days) for $\mu = 114$, $216$, and $444$, respectively. These estimates indicate that higher magnifications, which help shift the inferred source-frame masses below the mass-gap boundary, require increasingly massive lenses. While moderate magnifications can be produced by galaxy-scale lenses, the largest values would likely require very massive systems, such as galaxy clusters, which may reduce the astrophysical probability of such configurations. In addition, the long time delays predicted for the high-magnification scenarios, combined with the absence of identified lensed counterparts in current gravitational-wave catalogs~\cite{2025arXiv251216347T}, further reduce the plausibility of a strong-lensing interpretation. Taken together with the results above, this suggests that simple strong-lensing interpretations for GW190521 and GW231123 are disfavored under standard strong-lensing scenarios.

	\section{summary and discussion}\label{sec.5}\
    
    In this work, we investigate the interpretation of gravitational-wave (GW) signals from binary black hole (BBH) mergers located within the pair-instability supernova (PISN) mass gap, taking into account the effects of gravitational lensing. Using publicly available posterior samples from GWOSC, we adopt a posterior-rescaling approach to consistently propagate the full parameter uncertainties and explore how different lensing magnifications modify the inferred luminosity distances, redshifts, and source-frame masses of GW190521 and GW231123.

    We establish a direct mapping between the lensing magnification $\mu$ and the fraction of posterior samples with source-frame masses below the adopted mass-gap boundary of $65\,M_\odot$. Based on this relation, we define characteristic magnifications $\mu_{10}$, $\mu_{50}$, and $\mu_{90}$, corresponding to cases where $10\%$, $50\%$, and $90\%$ of the posterior samples fall below the mass-gap boundary, respectively. For GW190521, we obtain $\mu_{10}=3$, $\mu_{50}=9$, and $\mu_{90}=39$. At these magnifications, the corresponding source-frame masses of the primary black hole are $61^{+3}_{-6}\,M_\odot$, $56^{+8}_{-12}\,M_\odot$, and $42^{+19}_{-14}\,M_\odot$, with lensed redshifts of $z \sim 1.4$–$2.7$. These values are broadly consistent with previous lensing-based interpretations in the literature. For GW231123, we find significantly larger magnifications, with $\mu_{10}=114$, $\mu_{50}=216$, and $\mu_{90}=444$. Even at $\mu_{90}$, the source-frame mass of the primary black hole is reduced to $52^{+10}_{-10}\,M_\odot$, with a corresponding redshift of $z \sim 2.3$, indicating that extremely large magnifications are required to shift the majority of posterior samples below the mass-gap boundary.

    We further infer the corresponding lens properties under the singular isothermal sphere (SIS) model. For a representative configuration with $z_l = z_s/2$, the inferred lens masses for GW190521 range from $\sim 10^{10}\,M_\odot$ to $\sim 10^{13}\,M_\odot$, with associated time delays from days to months. In contrast, for GW231123, the required lens masses reach $\sim 10^{13}$–$10^{15}\,M_\odot$, corresponding to galaxy group or cluster-scale lenses, with time delays ranging from hundreds to thousands of days. These results indicate that achieving sufficiently large magnifications to move the inferred source-frame masses out of the mass gap requires increasingly massive lenses and correspondingly long time delays. In addition, we compute the lensing optical depth at the lensed source redshift and find that it remains small even at the highest magnifications considered (e.g., $\tau \sim 10^{-3}$ for GW231123 at $\mu_{90}$). Since the optical depth increases with redshift, the fact that it remains low even at the largest lensed redshift further disfavors a strong-lensing origin. Combined with the absence of identified lensed counterparts or repeated events in current gravitational-wave catalogs, and the long time delays predicted for high-magnification scenarios, these results suggest that simple strong-lensing interpretations for GW190521 and GW231123 are disfavored.

   We note that our posterior-rescaling approach is intended as a simplified and computationally efficient approximation. A full joint-inference analysis under the lensing hypothesis is more comprehensive, since it can consistently include the magnification, source parameters, population assumptions, and lensing priors, but it is also computationally more expensive. In contrast, our approach directly rescales the publicly available posterior samples and recomputes the corresponding source-frame masses and redshifts at fixed magnifications. The broad consistency between our rescaled results for GW190521 and the parameter ranges reported by LVK supports the use of this method as an effective first-order estimate. This comparison also motivates applying the same rescaling method to GW231123, for which we estimate the magnification required under the lensing hypothesis, and then combine this estimate with lens modeling and time-delay considerations to assess whether a simple strong-lensing interpretation is plausible. We also note that the SIS model used in this work is an idealized description of the lens mass distribution. Real lensing galaxies can deviate from spherical symmetry and may contain ellipticity, external shear, substructure, and other contributions~\cite{2010ARA&A..48...87T,2013A&A...559A..37S}. These effects can change the image magnifications and time delays, and may therefore affect the inferred lens mass and redshift required to produce a given magnification. Consequently, the lens properties derived from the SIS model should be interpreted as order-of-magnitude estimates rather than precise predictions for a realistic lens system. More flexible lens models, such as singular isothermal ellipsoids or power-law profiles with external shear, would be required for a more accurate assessment.
    
    During the current LVK observing runs, the detection of lensed BBH signals is expected to be extremely rare~\cite{2023MNRAS.526.3832J, 2024ApJ...970..191A}. However, with the operation of future third-generation detectors, the number of detected lensed events could reach hundreds~\cite{2015JCAP...12..006D,2022MNRAS.509.3772Y, 2023MNRAS.526..682G}, and a larger population of high-mass GW events is also expected. In this context, considering the lensing scenario in the inference of source-frame masses will provide an important complementary framework for interpreting BBH mergers that appear to reside within the PISN mass gap. 
    
    In summary, while gravitational lensing can in principle reduce the inferred source-frame masses of high-mass BBH mergers, our results indicate that the magnification required to reconcile GW190521 with the PISN mass gap is moderately large, whereas that required for GW231123 is astrophysically extreme. This, together with the corresponding lens properties and observational constraints, suggests that lensing alone is unlikely to fully explain these events under simple assumptions. Nevertheless, gravitational lensing remains an important effect in GW parameter inference and should be carefully considered in future analyses of high-mass BBH events.

\begin{acknowledgments}
We thank Mingqi Sun for useful discussion. You, Z.-Q. is supported by the National Natural Science Foundation of China (Grant No.~12305059); The Startup Research Fund of Henan Academy of Sciences (No.~241841224); The Scientific and Technological Research Project of Henan Academy of Science (No.~20252345003); Joint Fund of Henan Province Science and Technology R\&D Program (No.~235200810111); Henan Province High-Level Talent Internationalization Cultivation Program (No.~2024032).
\end{acknowledgments}


%

\end{document}